\documentclass{ws-ijmpd}

\begin{document}

\markboth{H. St\"ocker}
{Stable TeV - Black Hole Remnants at the LHC}

%
\catchline{}{}{}{}{}
%

\title{
Stable TeV - Black Hole Remnants at the LHC: \\
{\small Discovery through Di-Jet Suppression, Mono-Jet Emission and a Supersonic Boom in the Quark-Gluon Plasma }}

\author{H. St\"ocker}

\address{
FIAS- Frankfurt Institute for Advanced Studies\\ and\\
Institut f\"{u}r Theoretische Physik, Johann Wolfgang Goethe -
Universit\"{a}t,\\
 Max - von - Laue - Str.~1,\\
D--60438 Frankfurt am Main, Germany\\
E-mail: stoecker@FIAS.uni-frankfurt.de}

\maketitle

\begin{history}
\received{Day Month Year}
\revised{Day Month Year}
\comby{Managing Editor}
\end{history}

\begin{abstract}
The production of Large Extra Dimension (LXD) Black Holes
(BHs), with a new, fundamental mass scale of $M_f = 1$ TeV, has been
predicted to occur at the Large Hadron Collider, LHC, with the formidable
rate of \mbox{$10^8$} per year in p-p collisions at full energy, 14 TeV, and at
full luminosity.
\newline
We show that such LXD-BH formation will be experimentally observable at
the LHC by the complete disappearance of all very high $p_t$ (\mbox{$> 500$} GeV)
back-to-back correlated Di-Jets of total mass \mbox{$M > M_f = 1$}TeV, in the large
detectors ALICE, ATLAS and CMS.
\newline
We suggest to complement this clear cut-off signal at \mbox{$M > 2*500$} GeV in the
di-jet-correlation function by detecting the subsequent,  
Hawking-decay products of the LXD-BHs, namely either multiple high energy (\mbox{$> 100$}
GeV) SM Mono-Jets (i.e. away-side jet missing), sprayed off the
evaporating BHs isentropically into all directions or the thermalization of the multiple overlapping Hawking-radiation in a 
Heckler-Kapusta-Plasma: 
\newline
The extreme energy density of the Hawking Radiation may yield a Heckler-Kapusta-Hawking Quark-Gluon Plasma of SM - and SUSY -
particles at temperatures above the electroweak phase transition, which hydrodynamically (isentropically) evolves and cools until the
Quark-Hadron phasetransition and chemical freezeout at $T \sim 100$ MeV is reached.
Microcanonical quantum statistical calculations of the Hawking evaporation 
process for these LXD-BHs show that cold black hole remnants (BHRs) of Mass $\sim M_f$ remain leftover 
as the ashes of these spectacular Di-Jet-suppressed events. The BHRs are charged 
and can be detected as a track in the Central TPC of ALICE.
\newline
Strong Di-Jet suppression is also expected with Heavy Ion beams at the LHC, due to 
Quark-Gluon-Plasma induced jet attenuation at medium to low jet energies, $p_t$ \mbox{$< 200$} 
GeV. The (Mono-)Jets in these events can be used to trigger for Tsunami-emission of 
secondary compressed QCD-matter at well defined Mach-angles, both at the trigger side and at
the awayside (missing) jet. The Machshock-angles allow for a direct measurement of both the equation
of state EoS and the speed of sound $c_s$ via supersonic bang in the "big bang" matter. 
\newline
We discuss the importance of the underlying strong collective flow - the
gluon storm  - of the QCD- matter for the formation and evolution of these
Machshock cones. We predict a significant deformation of Mach
shocks from the gluon storm in central Au+Au collisions at RHIC and LHC
energies, as compared to the case of weakly coupled jets propagating through a static medium.
A possible complete stopping of $p_t>50$ GeV jets at the LHC in 2-3 fm yields nonlinear
high density Mach shocks in the quark gluon plasma, which 
can be studied in the complex emission and disintegration pattern of the possibly supercooled 
matter. We report on first full 3-dimensional fluid dynamical studies of the strong effects of a 
first order phase transition on the evolution and the Tsunami-like Mach shock emission of the QCD matter.
\end{abstract}


\section{Introduction}\label{intro}

The Frankfurt-born Astronomer Karl Schwarzschild discovered the first analytic solution of
the General Theory of Relativity \cite{Schwarzschild}. He layed the ground for studies of some of the most 
fascinating and un-understood objects in the universe, the Schwarzschild Black 
Holes. Recently it was conjectured that Black Holes (BHs) do also reach into the regime of
particle and collider physics: In the presence of additional
compactified large extra dimensions (LXDs), it seems possible to produce tiny black
holes at the LHC (Large Hadron Collider at the European Center for Nuclear Research, CERN, Geneva). 
This allows for tests of Planck scale physics and of the onset
of quantum gravity in the laboratory. Understanding black hole physics
is a key to the phenomenology of these new effects
beyond the Standard Model (SM).

The presence of additional spacelike dimensions \cite{Antoniadis:1990ew},
on top of our usual three space dimensions, seems ruled out by
experience, but these additional dimensions are compactified to small
radii, which explains why we have not yet noticed them.

During the last decade, several models \cite{add,rs1} using compactified LXDs as an
additional assumption to the quantum field theories of the Standard Model
(SM) have been proposed. The setup of these effective models is motivated
by String Theory, though the question whether our spacetime has additional
dimensions is well-founded on its own and worth the effort of examination.

The models with LXDs provide us with a useful description to predict
first effects beyond the SM. They do not claim to be first principles
theories. Instead, their simplified framework allows the derivation of
testable results, which can in turn help us to gain insights about the
underlying
theory.

Large extra dimensions have e.g. been incorporated into the framework of the
SM by the {\sc ADD}-model proposed by Arkani-Hamed, Dimopoulos and Dvali
\cite{add}, who add $d$ extra spacelike dimensions without curvature, in general each of them
compactified to the same radius $R$. All SM particles are confined to our 3+1
brane, while gravitons are allowed to propagate freely in the 3+d+1 dimensional bulk.
In the following, we consider the phenomenological consequences of the ADD-
model \cite{add} with a new fundamental mass-scale  $M_{\rm f}$. 
The new and the apparent Planck scales are related by
\begin{eqnarray}
m_{\rm p}^2 = M_{\rm f}^{d+2} R^d. \quad \label{Master}
\end{eqnarray}

The radius $R$ of these extra dimensions, for $M_{\rm f}\sim$~TeV, can be
estimated with Eq.(\ref{Master}) and typically lies in the range from $10^{-1}$~mm to 
$\sim 100$ fm for $d$ from $2$ to $7$. Therefore, the inverse compactification radius
$1/R$ lies in energy range eV to MeV, respectively. The case $d=1$ is
excluded. It would result in an extra dimension about the size of the solar system. 
d=2 has recently been excluded by Cavendish-type experiments.

\section{Estimate of LXD- Black Hole formation crosssections at the LHC}

The most exciting signature of {\sc LXD}s is the possibility of black hole
production as discussed in the present note 
\cite{dim}$^-$\cite{Hossenfelder:2005bd}
and in ultra high energetic cosmic ray events
\cite{cosmicrayskk,cosmicraysbh}. For recent updates on constraints on the
parameters $d$ and $M_f$ see e.g. \cite{Cheung:2004ab}.
In the standard $3+1$ dimensional space-time, the production of black holes requires a
concentration of energy-density which can not be reached in the
laboratory, while in the higher dimensional space-time, gravity becomes
stronger at small distances, and, therefore, the event horizon is located at a larger radius, 
$R_H\sim 10^{-4} fm$.

We are interested in the case where the black hole has a mass close to the new fundamental scale 
$M_{BH}\approx M_f\approx 1$ TeV. This corresponds to a radius $R_S=R_{BH}=R_H\approx 10^{-4}fm$ 
close to the inverse new fundamental scale $M_f$, and thus $R_H \ll R$.

Higher dimensional Schwarzschild-metrics have been derived in \cite{my}
with the Schwarz\-schild\-radii $R_H$
\begin{equation} \label{ssradD}
R_H^{d+1}=
\frac{2}{d+1}\left(\frac{1}{M_{\rm f}}\right)^{d+1} \; \frac{M}{M_{\rm f}}.
\quad 
\end{equation}

Naturally, black holes with mass of about the new fundamental mass $M_{BH} \sim M_{\rm f}$, have 
Schwarzschild radii of about the new fundamental length scale $L_{\rm f}=1/M_{\rm f}$ (which justifies the use 
of the limit $R_H\ll R$).

As for $M_{\rm f}\sim$~1TeV this radius is $R\sim 10^{-4}$~fm, p-p-collisions
of $\sqrt{s}=14$ TeV at the LHC, which allow for very hard (say: $\Delta$p = 1 TeV)
parton-parton two-body scattering events, (Di-Jets of $E^{Tot}_{Dijet}\sim 1$ TeV) will, due
to the uncertainty relation, yield two partons with impact parameters closer together than
$\Delta x = \frac{1}{10000}$ fm. This corresponds to the Schwarzschild radius of the
two partons with M $> 1$ TeV, close to the new fundamental scale $M_{\rm f}\sim 1$~TeV - 
black holes can be created at the LHC in the ADD model!

The LXD-black hole production cross section can be approximated by the classical geometric 
cross-section
\begin{eqnarray} \label{cross}
\sigma(M)\approx \pi R_H^2, \quad
\end{eqnarray}
which only contains the fundamental Planck scale as coupling constant. This classical cross 
section is under debate \cite{Voloshin:2001fe,Rychkov:2004sf}, but seems justified at least up 
to energies of $\approx 10 M_{\rm f}$ \cite{Solodukhin:2002ui}.

Semi classical considerations yield form factors of order one\cite{Formfactors}, which take into 
account the fact that not the whole initial energy can be captured behind the Schwarzschild 
horizon. The naive classical result remains valid also in string theory \cite{Polchi}.

Angular momentum $J  \approx 1/2 M b$ considerations change the results by a factor $2$ \cite{Solo}.
The black hole also carries charge, which gives rise to the exciting possibility of naked 
singularities \cite{Casadio:2001wh}.

What is the threshold for the black hole formation? From Thorn's General Relativistic arguments,
two point like particles in a head on collision with zero impact parameter will {\sl always} 
form a black hole, no matter how high or low their energy. At low energies, however, this 
overlap is improbable due to the spread of the wave functions by the uncertainty relation. 
This results in a necessary minimal energy to allow for the required
close approach. Also this threshold is of order $M_{\rm f}$, though the
exact value is unknown - quantum gravity effects should play an important role for the wave 
functions of the colliding particles.

Setting $M_{\rm f}\sim 1$TeV and $d=2 - 7$ one finds $\sigma \sim 400$~pb $- 10$~nb.
Using the geometrical cross section formula, it is now possible to compute
the differential cross section ${\mathrm d}\sigma/{\mathrm d}M$  which is given by summation 
over all possible parton interactions and integration over the momentum fractions, where 
the kinematic relation $x_1 x_2 s=\hat{s}=M^2$ has to be fulfilled. This yields
\begin{eqnarray} \label{partcross}
\frac{{\rm d}\sigma}{{\rm d}M}
&=&  \sum_{A_1, B_2} \int_{0}^{1} {\rm d} x_1 \frac{2 \sqrt{\hat{s}}}{x_1s}
f_A(x_1,\hat{s})
f_B (x_2,\hat{s})  \sigma(M,d).   \quad
\end{eqnarray}

\begin{figure*}[htb!]
\centerline{\includegraphics[width=0.7\textwidth]{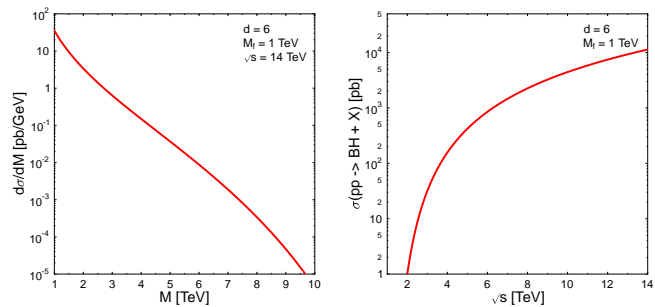}}
\caption{The left plot shows the differential cross section for black hole
production in proton-proton-collisions at the {\sc LHC} for $M_{\rm f}=1$~TeV. 
The right plot shows the integrated total cross section as a function of the 
collision energy $\sqrt{s}$. In both cases, the curves for various
$d$ differ from the above depicted ones by less than a factor
10.\protect\cite{Hossenfelder:2005bd}}
\protect{\label{dsdm}}
\end{figure*}

A numerical evaluation \cite{Hossenfelder:2005bd} using the {\sc CTEQ} -
tables results in the differential cross section displayed in Figure \ref{dsdm}, left. 
Most of the black holes created have masses close to the production threshold. This is due to the fact
that at high collision energies, or small distances, respectively, the
proton contains a high number of small x, low
energy gluons and the total energy is distributed among them.

It is now straightforward to compute the total BH- cross section by
integration over Eq. (\ref{partcross}), see Figure \ref{dsdm}, which yields a production cross section of
about 10 nanobarn. $N_{\rm BH}=10^9$ black holes may be created at the LHC per year with the estimated full LHC
luminosity $L=10^{34}{\rm cm}^{-2}{\rm s}^{-1}$ at $\sqrt{s}=14$~TeV: About one black hole per 
second would be created \cite{dim}.

LXD- Black Hole production would have dramatic consequences for future
collider physics: Once the collision energy crosses the threshold for
black hole production, no further information about the structure of
matter at small scales can be extracted - this would be ''{\sl the end of
short distance physics}''\cite{Giddings3}.

\section {Suppression of high mass correlated Di-Jet Signals -  LXD-Black Hole Formation at the LHC}

The above findings led to a high number of publications on the topic of
TeV-mass black holes at colliders 
\cite{dim,ehm,Giddings3,Hofmann:2001pz,Hossenfelder:2001dn,Chamblin:2002ad,Anchordoqui:2002cp}, for 
hadronic collisions as well as for heavy ion collisions \cite{own2}:

Per PbPb event, the number of BHs is increased (compared to pp) more than thousandfold as there are about 200-400
p-n collisions at a central ($b\sim 3$~fm) impact parameter event, which occurs with 400 mb cross section.

But how can we observe with certainty such a rare, exotic process, in the
enormous background of a billion p-p events or so?

Are there unambigous BH-observables, so robust and unique that an
experimental signal is achievable? {\bf Yes: Di-Jets vanish above $M_f$! Therefore:}
\begin{eqnarray}
~&&\mbox{IF (LXD-Black Holes at 14 TeV in pp @ LHC)}
\nonumber\\~&&\mbox{.THEN.}\nonumber\\
~&&\mbox{(No High pT-events, no 2 $\cdot$ 500 GeV Di-Jets @ LHC)}
\end{eqnarray}

The first, cleanest signal for LXD-BH- formation at the LHC is the
complete suppression of high energy back-to-back- correlated Di-Jets with
$M > M_f$: those two very high energy partons, $E_{jet}\approx$ one-half $M_f$ each, 
i.e. $p_t \geq 500$ GeV each, which usually define the Di-Jets in the standard model, now end up
inside the black hole, instead of being observable in the detector! Di-Jets with $E_{Dijet} > M_f$
can not be emitted! A clean signature for BHs, indeed: The end of short distance 
physics.

The threshold cut-off due to black hole production, at $M \leq$ 1 TeV or so in
the transverse momentum spectra, also seems ro lead to a decrease in the single particle spectra 
at higher pt-values, above $p_t$ = 1 TeV\cite{Chamblin:2002ad,Lonnblad:2005ah}.

However, the high $p_t$ - single particle spectra at high $M$ may be filled up again due to Hawking 
radiation Mono-Jets, as discussed in the next paragraph. Therefore, the Di-Jet Suppression proposed 
here as signals for LXD-BHs, is the preferred observable.

In analogy to Heckler and Kapusta (3+1-dim BHs), Anchordoqui and Goldberg \cite{Anchordoqui:2002cp} show that, for LXD-BH- 
Hawking radiation, emitted partons are closely spaced outside the Schwarzschild horizon. Hence, 
the partons do not fragment into hadrons, as strings in vacuum would, but they form a quark-gluon- 
(plus lepton- \& EW-gauge vector boson-) plasma.

Above the temperature of the electroweak phase transition, i.e. at hundred GeV, even EW-particles melt 
in the superhot medium - like rhos in the QCD plasma - and can emit dilepton pairs at the Z's in-medium mass.

Final thermal hadron freeze out then occurs only after isentropic hydrodynamical flow, at much lower temperatures, 
far from the horizon, i.e. at temperatures of the QCD phase transition scale, T=200 MeV.

The energy spectrum of the particles emerging from the "chromosphere" is found to be relatively soft: hard hadronic 
jets are almost entirely suppressed. The Jets are replaced by an isotropic distribution of soft photons and hadrons, 
with hundreds of particles in the few GeV range. This distinctive signature for black hole events at LHC should easily 
be discovered in the ALICE TPC. 

\section{Hard, Isotropic Multiple Monojet Emission as Signal for Hot LXD-
Black Hole Hawking- Evaporation}

Once produced, the black holes may undergo an evaporation process
\cite{Hawk1}
whose thermal
properties carry information about the parameters $M_{\rm f}$ and $d$. An
analysis of the evaporation
will therefore offer the possibility to extract knowledge about the
topology of our
space time and the underlying theory.

The evaporation process can be categorized in three characteristic stages
\cite{Giddings3,Hossenfelder:2001dn,Hossenfelder:2005ku}:

\begin{enumerate}
\item {\sc Balding phase:} In this phase the black hole radiates away the
multipole
moments it has inherited
from the initial configuration, and settles down in
a hairless state. During this stage, a certain fraction of the initial
mass will be
lost in gravitational
radiation.
\item {\sc Evaporation phase:} The evaporation phase starts with a spin
down phase
in which the Hawking
radiation carries away the angular momentum, after which it proceeds with
emission
of thermally
distributed quanta until the black hole reaches Planck mass. The radiation
spectrum
contains
all SUSY- and Standard Model particles, which are emitted on our brane, as well as
gravitons,
which are also
emitted into the extra dimensions. It is expected that most of the initial
energy is emitted in during this phase in Standard Model particles.
\item {\sc Planck phase:} Once the black hole has reached a mass close to
the Planck
mass, it falls into
the regime of quantum gravity and predictions become increasingly
difficult. It is
generally
assumed that the black hole will then either completely decay in a few
Standard
Model particles or
a stable (charged or neutral) Black Hole Remnant (BHR) will be left, which carries away the remaining energy and mass.
\end{enumerate}

To understand the propsed signature caused by black hole production,
namely the occurence of multiple Monojets, in those Di-Jet suppressed
events, we have to examine the Hawking- evaporation process in detail.
The evaporation rate ${\rm d}M/{\rm d}t$ also in higher dimensional
space-times can be
computed using the thermodynamics of black holes. One finds for the
BH-Hawking-like temperature the
relation
\begin{eqnarray} \label{tempD}
T=\frac{1+d}{4 \pi}\frac{1}{R_H} \quad,
\end{eqnarray}
where $R_H$ is a function of $M$ by Eq. (\ref{ssradD}). The smaller the
black hole,
the larger is its temperature.

Hence, the very hard radiation of the exploding, tiny black holes is a
second signature.

Typical temperatures at the end of the lifetime are several hundred GeV. Since most
of the particles of the black body radiation are emitted with $\sim 100$~GeV average
energy, we can estimate the
total number of emitted particles to be of order $10$-$100$.
This high temperature results in a very short lifetime such that the black
hole will decay close to the primary interaction region and can be interpreted as
a metastable intermediate state.

\begin{figure*}[ht!]
\centerline{\includegraphics[width=0.7\textwidth]{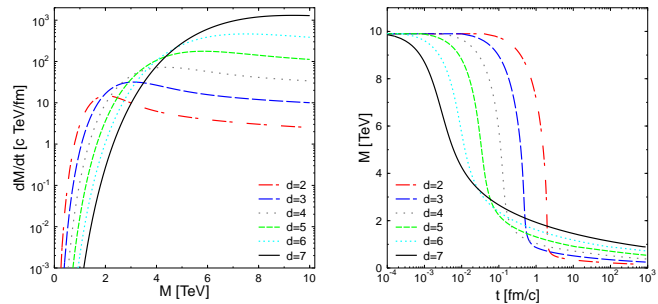}}
\caption{The left plot shows the black hole's evaporation rate in as
a function of the initial mass for various $d$. The right plot shows the
time evolution of the mass of a black hole with an initial mass of 10~TeV. \label{mdot}}
\end{figure*}

Integrating the thermodynamic identity d$S/$d$M = 1/T$ over $M$ yields 
the entropy
\begin{eqnarray}
S(M)
&=& 2 \pi  \frac{d+1}{d+2} \left( M_{\rm f} R_H \right)^{d+2}\quad.
\end{eqnarray}
For the number density of the states one has to take into account that for
the
typical collider-produced black hole. The emission of one particle will have a non-negligible
influence on
the total energy of the black hole. This problem can appropriately be
adressed by
including the back-reaction of the emitted quanta as has been
derived in \cite{Page,backreaction}. It is found that in the regime of
interest
here, when
$M$ is of order $M_{\rm f}$, the emission
rate for a single particle microstate is modified and given by the change
of the black
hole's entropy
\begin{equation} \label{nsingle}
n(\omega) =  \frac{\exp[S(M-\omega)]}{\exp[S(M)]}\quad.
\end{equation}
For Boltzmann-statistic this leads then to the spectral energy density
\begin{eqnarray}
\varepsilon = \frac{\Omega_{(d+3)}}{(2 \pi)^{3+d}}
{\mathrm e}^{-S(M)} \sum_{j=1}^{\infty} \frac{1}{j^{d+4}} \int_0^{M} {\mathrm
e}^{S(x)} (M-x)^{3+d} {\mathrm d}x
\quad,
\end{eqnarray}
where the value of the sum is given by a $\zeta$-function. From this we
obtain the
evaporation rate
\begin{eqnarray} \label{mdoteq}
\frac{{\mathrm d}M}{{\mathrm d}t} = \frac{\Omega_{(d+3)}^2}{(2\pi)^{d+3}}
R_H^{2+d}
\zeta(4+d) \;
{\rm e}^{-S(M)}
\int_{0}^{M} (M-x)^{(3+d)} {\rm e}^{S(x)} {\mathrm d}x \quad.
\end{eqnarray}
A plot of this quantity vs. $M$ for various $d$ is shown in Figure \ref{mdot},
left. The 
time-dependence of the mass 
is shown in Figure \ref{mdot}, right. We see that the evaporation process
of $10$~TeV BHs slows down
in the late stages and
enhances the lifetime of the black hole \cite{Hossenfelder:2001dn}.
The predominantly produced "light" primary BHs, $m_{BH}^{initial}\approx 1$~TeV, exhibit larger lifetimes. 
In case of stable Black Hole Remnants (BHRs), the lifetime will approach infinity - they will behave as a new kind of
elementary particle with $M_{BHR}\sim M_f$, which can absorb mass (e.g. protons), but evaporates that mass swiftly in
form of photons and lepton pairs and decays then back into the BHR-groundstate.

To perform a realistic simulation of the evaporation process, one has to
take into account all Standard Model particles. In the extra dimensional
scenario, SM particles are bound to our submanifold, whereas the gravitons
are allowed to enter all dimensions. It has been argued
that black holes emit mainly on the brane \cite{ehm}.
A very thorough description of these evaporation characteristics has been
given in
\cite{Kanti:2004nr}
which confirms the expectation that the bulk/brane evaporation rate is of
comparable magnitude but the brane modes dominate.

For recent reviews on TeV-scale black holes see also
\cite{Landsberg:2002sa} and references therein. Several experimental
groups have included LXD- BH searches into their research programs for
physics beyond the Standard Model, in particular the ALICE-, ATLAS- and CMS-
Collaborations at the LHC\cite{Humanic}. Both the PYTHIA 6.2 \cite{PYTHIA} and
the CHARYBDIS \cite{CHARYBDIS} event generators allow for a simulation of
black hole events and data reconstruction from the
decay products. Such analysis has been summarized in Ref. \cite{Atlas} and
Ref. \cite{BHex}, respectively.

Ideally, the energy distribution of the decay products allows for a
determination of the temperature (by fitting the energy spectrum to the
predicted shape) as well as of the total mass of the BH (by summing up all
energies).
This then will allow for a reconstruction of the scale $M_{\rm f}$ and the
number of extra dimensions.

Due to the high energy captured inside the black hole, its decay will be
spectacular. Its distinct features allow for a second, independend
signature, which can be used for the BH detection: in the very same event
sample, where high energy Di-Jets are suppressed, a very
high multiplicity of high energy Mono-Jets, much higher multiplication than in SM
processes can be observed.

Furthermore, the thermally evaporating black hole yields a nearly
isotropic decay pattern, with  a high sphericity of the event.

\section{Heckler-Kapusta-Hawking SM+SUSY-Plasma - above the electroweak phasetransition, formed in pp $\rightarrow$ BH
reactions at the LHC}

The energy density of the multiple Hawking Mono-Jets emitted from the evaporating BHs and BHRs produced in pp-collisions
is enormous: Several TeV are emitted within a 4-Sphere of $\sim 10^{-12} \hbox{fm}^4$/c, implying energy densities of
$\geq 10^9\, \hbox{TeV fm}^{-3}$, i.e. many orders of magnitude higher than the energy densities expected for the Quark-Gluon
Plasma $e\sim 500 \,\hbox{GeV fm}^{-3}$ to be created in Pb-Pb - collisions at the LHC at $\sqrt{s}=5.5$~ATeV.

Hence the question arises whether - at this enormous energy density - the multiple jets thermalize to form a
ultra-hot $T\gg T_{EW}\gg T_{QCD}$ Plasma of Standard-Model plus SUSY-particles. 

Such a "hot Heckler-Kapusta-Hawking Plasma" scenario has been studied intensely for primordial 3-1-dimensional Black 
Holes in old cosmic radiation components by Heckler-Kapusta and coworkers \cite{Kapusta:2000xt,Heckler:1997jv,Heckler:1995qq}. 
For the LXD-BHs and BHRs to be studied at the LHC 
a similar process might happen: The HKH-plasma for $T\sim 1$~TeV should contain many "massless" SM Paricles; as $T$ is 
above the electroweak phasetransition temperature, the bare masses of e.g. $W^{+/-}$ and $Z$ as well as light 
supersymmetric partners may become accessible to experiment. The HKH-plasma and the multiple-mono-jet emission may be 
distinguished by Hanbury Brown Twiss (HBT-) two- and more particle correlation measurements of the source size and 
life time of the system.

Interesting questions to study are the properties of SM- and SUSY-particles (masses, width) in the hot $(T\gg T_{Ew})$
medium, thermalization and viscosity of this SM-SUSY "state" of matter, hydrodynamic expansion, abundant emission of
(otherwise rare) quarks $(b,t)$ and leptons.

\section{High Intensity Beams of Low Energy Protons as a potent Energy Source through Hawking-Radiation of Black 
Hole Remnants (BHRs)}

Stable remnants could be used as catalysors to capture and convert, in accord with  $E=mc^2$, high intensity beams of low 
energy baryons (p,n, nuclei), of mass $\sim 1$ AGeV, into photonic, leptonic and light mesonic Hawking radiation, thus serving 
as a source of energy with 90\% efficiency (as only neutrinos and gravitons would escape the detector/reactor).

If BHRs (Stable Remnants) are made available by the LHC or the NLC and can be used to convert mass in energy, then the 
total 2050 yearly world energy consumption of roughly $10^{21}$ Joule can be covered by just $\sim 10$ tons of abitrary material, 
converted to radiation by the Hawking process via $m = E/c^2 = 10^{21} \hbox{J}/(3\cdot 10^8 \hbox{m/s})^2 = 10^4$ kg 
\cite{HorstPatent}. 

\section{Formation of stable Black Hole Remnants and Single Track Detection in the ALICE-TPC}

Numerical simulations have so far assumed mostly that the black holes decay completely
into SM- particles. From a theoretical point of
view, however, there are strong indications that the black holes do not
evaporate completely, but rather leave a meta-stable black hole remnant,
dubbed Black Hole Remnant (BHR) or relic\cite{Ben,Marcus} - 
do these relics leave the detector? Do they still
emitt radiation?

\begin{figure}[hpt!]
\centerline{\epsfxsize=10cm\epsfbox{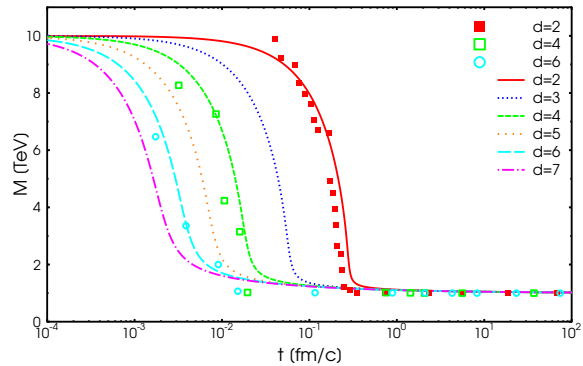}}
\caption{The mass evolution for a black hole of initial mass $M=10$~TeV and
various $d$. Here, we set $M_{\rm R} = M_{\rm f} =1$~TeV.}
\label{mtrel2}
\end{figure}

Fig.~\ref{mtrel2} shows that this is not the case: the mass evolution of
the  produced black holes stabilizes rapidly, $t < 1$fm/c, the average
energy of emitted particles drops to zero within $10$ fm.

The numerical results obtained (using the black hole event
generator CHARYBDIS and the observables computed within the PHYTHIA
environment) agree very well with the analytically computed results.

The fast convergence of the black hole mass is due to the fact that
the spectral energy density has a maximum at energies $\sim 3 T$. If the
mass of the black hole decreases, emission of high energy particles is
no longer possible because of energy conservation.

\begin{figure}[hpt!]
\centerline{\epsfxsize=10cm\epsfbox{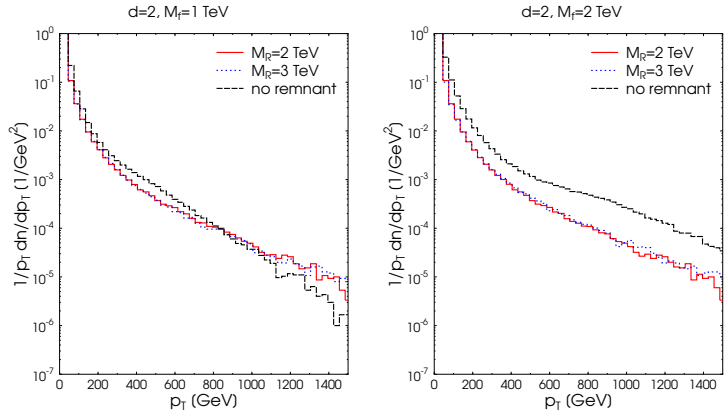}}
\caption{Transverse momentum distribution after fragmentation with
final (two-body) decay in contrast to the formation of a black hole
remnant. \protect\cite{Koch:2005ks}}
\label{ptd2}
\end{figure}
Figure \ref{ptd2} shows the $p_T$-spectrum after fragmentation 
as predicted by the CHARYBDIS-code.
One clearly sees the additional contribution from the final decay, which causes
a bump in the spectrum. This bump is absent in the case of remnant
formation. This graph does not include background, but the ALICE detector can differentiate 
LXD-BHs from QCD-Background \cite{Cortese:2002kf}. Moreover, the microcanonical 
Hawking-Evaporation produces Multiple Mono-Jets even when BHRs are assumed. 
The BHR-remnant signal in p(14 TeV)p $@$ ALICE is clearly distinguishable from
disappearing BHs!

\begin{figure}[hpt!]
\centerline{\epsfxsize=10cm\epsfbox{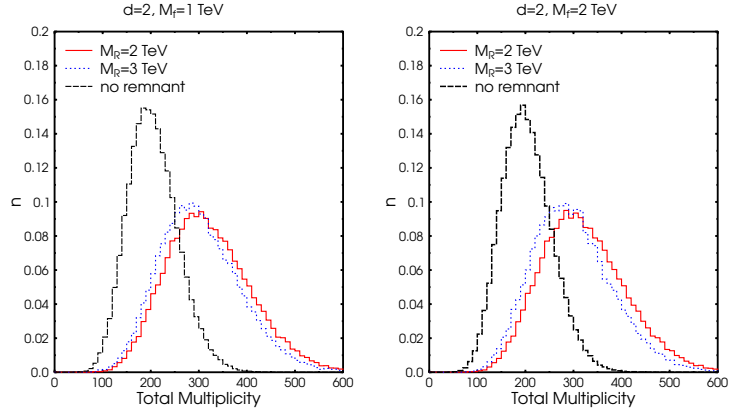}}
\caption{Total multiplicity with final (two-body) decay and with formation
of a black hole remnant for $d=2$. \protect\cite{Koch:2005ks}}
\label{multd2}
\end{figure}

Figure \ref{multd2} displays the total multiplicities in such an event: when a
black hole remnant is formed, the total multiplicity in CHARYBDIS is increased 
in spite of the fact that less energy is available (due to the missing BHR-mass) due to the
many additional low energetic particles that are emitted in the late stages,
instead of a final decay with $2 - 5$ hard particles.

Note that this multiplicity increase is not an effect of the remnant
formation itself, but rather it stems from the treatment of the decay in
the microcanonical ensemble used in
the present calculation: the black hole remant, BHR, evaporates more particles
with lower energy. The effect of hanced production of moderate energy secondaries will be stronger even if the Multi-Jets thermalize and
form a QGP which then expands and cools. 

Charged Black Hole Remnants should be observable in the TPC of the ALICE-Detector at the LHC as a magnetically very
stiff single - or double charged track of very high momentum, large mass, moderate velocity.

\section{Di-Jet-suppression in Relativistic Heavy Ion Collisions: 
formation of dense Quark-Gluon Matter at RHIC and LHC?}

Remarkably strong bulk elliptic flow patterns of
Dense Quark Matter have been observed at the Relativistic Heavy Ion Collider RHIC at BNL, N.Y.
Even stronger collective effects are expected in Heavy Ion Collision. Apparantly
the pressure of the matter formed at RHIC is very high, $P_{QCD} >> 1 \hbox{GeV/fm}^3$!
The elliptic $V_2$-Flow is THE Barometer  for the Equation of State.

Di- Jet suppression have been observed - about 50\% can be attributed to Hadron Rescattering \cite{Cassing:2004xr}.
pQCD shows high opacity of QCD-matter to Di-Jets: only Monojets are emitted!

Are Mach-Cones or Wakes caused by Di-Jets in QGP MATTER observed!?
The speed of sounc, $c_s$, is the response to jet-penetration probes  - weaks can lead the measurement of 
dispersion relations. 
Transport coefficients (viscosity, conductivity) can be determined.

At RHIC huge $\langle v_2\rangle$ of high $p_t$ particles is observed!
The QG-Plasma Pressure is much larger than the Hadron Pressure !
Therefor, QUARK MATTER FLOWS! QGP is early-on
thermalized - and can create huge Plasma pressure!

Ultra-dense QCD matter can be discovered with heavy ions in the LHC -just like at
RHIC- through three steps:
\begin{enumerate}
\item[1.] $P_{QCD}$: remarkable bulk elliptic flow patterns, $v_2>30\%$
\item[2.] pQCD: higher opacity to Di-Jets, tomography even for high $p_t$-jets
\item[3.] $c_s$: response to penetrating probes, Mach-Cones and Tsunami emission.
\end{enumerate}

Parton energy loss studies in pQCD show that the dominant mechanism is
gluon bremsstrahlung after collisions in the medium. The scattering power of the
QCD medium is proportional to the "gluon density" and hence to the
gluonic $T_{\mu\nu}$. 
But up to now $3<=>2$ gluon collisions have not yet been incorporated, however, see Xu and Greiner \cite{Xu}!
Therefore, for Jet Quenching and Tomography of Quark Gluon Matter at RHIC:
(Gyulassy, Vitev, Wang \cite{Gyulassy:2003mc}) one needs very large gluon densities!
Jet Quenching in medium has been discovered at RHIC, but not understood.

How much jet quenching is due to Hadrons FSI ?
Jets interact in both, Plasma- and Hadron Phase!

Hadronic rescattering - responsible for $50\%$ energy loss?\cite{Cassing:2004xr} 

An additional 50\% of $p_T$ -suppression is due to hadron rescattering: the
expansion of small color transparency  configuration yields  50\% hadronic
quenching, an additional $\sim 50\%$ QGP is needed.

Also here, half of the effect is due to hadronic, half due to partonic interaction\cite{Cassing:2004xr}.
How much of the suppression of Di-hadron correlations is due to hadronic FSI (final state interactions)?

\section{Mach Shock Cones and nonlinear Tsunamis induced by Jets Stopping
in Quark-Gluon Plasma}

Sideward peaks have been recently
observed\cite{Adl03a,Adl03b,Wan04,Jac05,Bue05} in azimuthal distributions of
secondaries associated with a high~$p_T$ hadron in central Cu+Cu and Au+Au
collisions at \mbox{$\sqrt{s_{NN}}=62$ and 200 GeV}. Experimental
data of the STAR and PHENIX collaborations both show that the
''away--side'' dip
of two--hadron correlations at $\Delta\phi=\pi$ disappears for peripheral
events
and with raising the transverse momentum of the associated particle.

In Ref.\cite{Sto04} such
peaks were predicted as a signature of Mach shocks created by partonic
jets propagating through a quark--gluon plasma (QGP) formed in a
heavy--ion collision. Analogous Mach shock waves were studied
previously in cold nuclei \mbox{\cite{Hof74,Sto80,Sto86,Cha86,Ris90}}.
This phenomenon has subsequently been studied in a numerical
Ref.\cite{Cas04} linearized fluid-dynamical approach.

Other possible explanations of the sideward
peaks in the Mono-jet distributions have been offered \cite{dremlin06}. For example, it is
argued in Refs.\cite{Sto04,Rup05} that Mach-like motions of
quark--gluon matter can appear via the excitation of collective
plasmon waves, "wakes", by the moving color charge associated with the
leading jet.

A similar mechanism, electron emission induced by heavy-ion
irradiation of metal foils, was predicted in Ref.\cite{Sch78}.
Later on this effect was observed experimentally\cite{Fri80}.

Keep in mind that the colored plasmon waves can be produced
only in a nonideal QGP\cite{Rup05}. According to
Refs.\cite{Dre79,Maj05}, excitation of transverse plasmon waves
in QGP may lead to a conical, Cherenkov-like emission of particles.
However, this is possible only for a strongly coupled QGP.

A high energy parton moving
in the storm of moving quark-gluon matter deposits a fraction of the
parton's energy
and momentum along its trajectory. Colorless sound waves then produce the
Mach region of perturbed collective flow
behind the leading particle.

In the fluid rest frame (FRF) the Mach region
has a conical shape with an opening angle (with respect to the direction
of particle propagation) given by the expression
\begin{eqnarray}\label{mac1}
\theta_M=\sin^{-1}\left(\frac{c_s}{\widetilde{v}}\right)\,,
\end{eqnarray}
where $c_s$ denotes the sound velocity of the unperturbed
fluid and $\widetilde{\bm{v}}$ is the particle velocity with respect to the
fluid.

This Mach-Cone can be formed only if $\widetilde{v}> c_s$.
Strictly speaking, Eq. (\ref{mac1}) is applicable only for weak,
sound-like linear perturbations.

Following Refs.\cite{Sto04,Cas04}
one can estimate the angle of preferential emission of partonic or
hadronic secondaries associated with a fast jet in the QGP.

The appearance of the Mach shock leads to angular maxima in the particle
emission, corresponding to \mbox{$|\Delta\phi-\pi|=\pi/2-\theta_M$}.
In the case of ideal QGP ($c_s=1/\sqrt{3}$),
substituting $\widetilde{v}\simeq 1$ into Eq~(\ref{mac1}), one obtains that
$\Delta\phi\simeq\pi\pm 0.96$.

This agrees well with positions of maxima
of the two-particle distributions observed in heavy--ion collisions
at RHIC energies. Therefore, by measuring the two--particle
angular distributions, one should be able to easily extract the sound
velocity of the QGP.

These estimates correspond to the idealized case of
homogeneous, infinite, static matter.

However, the systems are finite, and collective expansion flow is known to
be strong in relativistic collisions of heavy
nuclei\cite{Nux04}.

For example, thermal fits of RHIC data give for
most central events the average radial flow velocities $u_f\sim 0.6$\,.
This value is as large as $c_s$ in the deconfined phase!

Due to the hadronization effects, $c_s$
will be strongly time--dependent. Characteristics of Mach shocks in
expanding quark--gluon
matter have been considered in Refs.\cite{Sat05,Ren05}

Typical flow parameters at RHIC and LHC energies yield \cite{Sat05} shapes
and orientation of Mach regions which are strongly modified
as compared to the case of static medium.

Especially strong deformation of Mach-Cones takes place when the flow
velocity is orthogonal to the direction of the jet's propagation. In this
case the collective flow acts like a storm, deflecting the Mach-Cone
in the direction transversal to the di-jet axis.

As a result, the shape of Mach region becomes asymmetric with respect
to the jet trajectory in the global center of mass frame (CMF).

\begin{figure}[hpt!]
\vspace*{-6cm}
\centerline{\epsfxsize=10cm\epsfbox{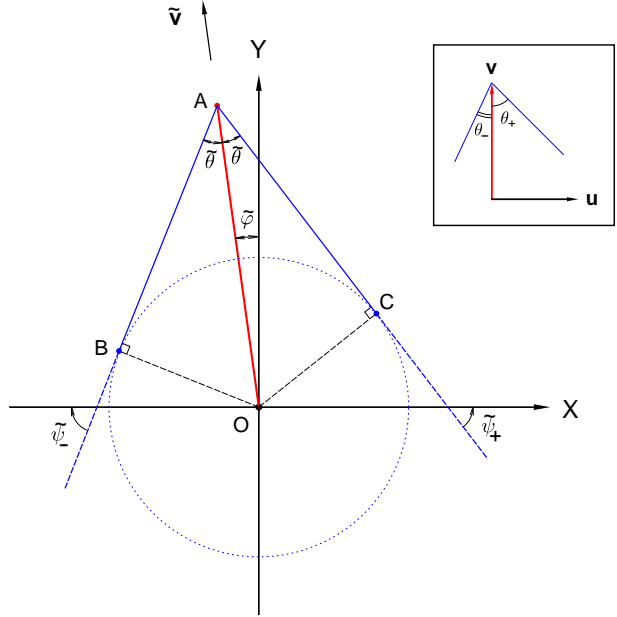}}
\caption{Mach region created by jet
moving with velocity $\bm{v}$ orthogonal to the fluid velocity~$\bm{u}$\,.
The main plot and the insert correspond to FRF and CMF, respectively.
It is assumed that jet moves from $O$ to $A$ in FRF. The dotted
circle represents the front of sound wave generated at point $O$.}
\label{mach-fig1}
\end{figure}

This is illustrated in Fig.~\ref{mach-fig1}. The insert
shows that the boundaries of Mach region have
different angles, $\theta_+\neq\theta_-$, with respect to the jet velocity
$\bm{v}$ in the CMF.
By using the Lorentz transformation from CMF to FRF, the following
formulae is obtained for $\theta_\pm$ in the weak shock approximation:
\begin{eqnarray}\label{mac2}
\tan{\theta_\pm}\simeq\gamma_u
\frac{ \gamma_s\, c_s\pm\gamma_u\, u}
{ 1\mp\gamma_s\, c_s\gamma_u, u}\,,
\end{eqnarray}
where $\gamma_u=(1-u^2)^{-1/2}, \gamma_s=(1-c_s^2)^{-1/2}$\,.
One can see that for small flow velocities the difference of the
Mach angles $\theta_\pm$ in moving and static matter is approximately
linear in $u$\,.

\begin{figure}[hpt!]
\vspace*{-6cm}
\centerline{\epsfxsize=10cm\epsfbox{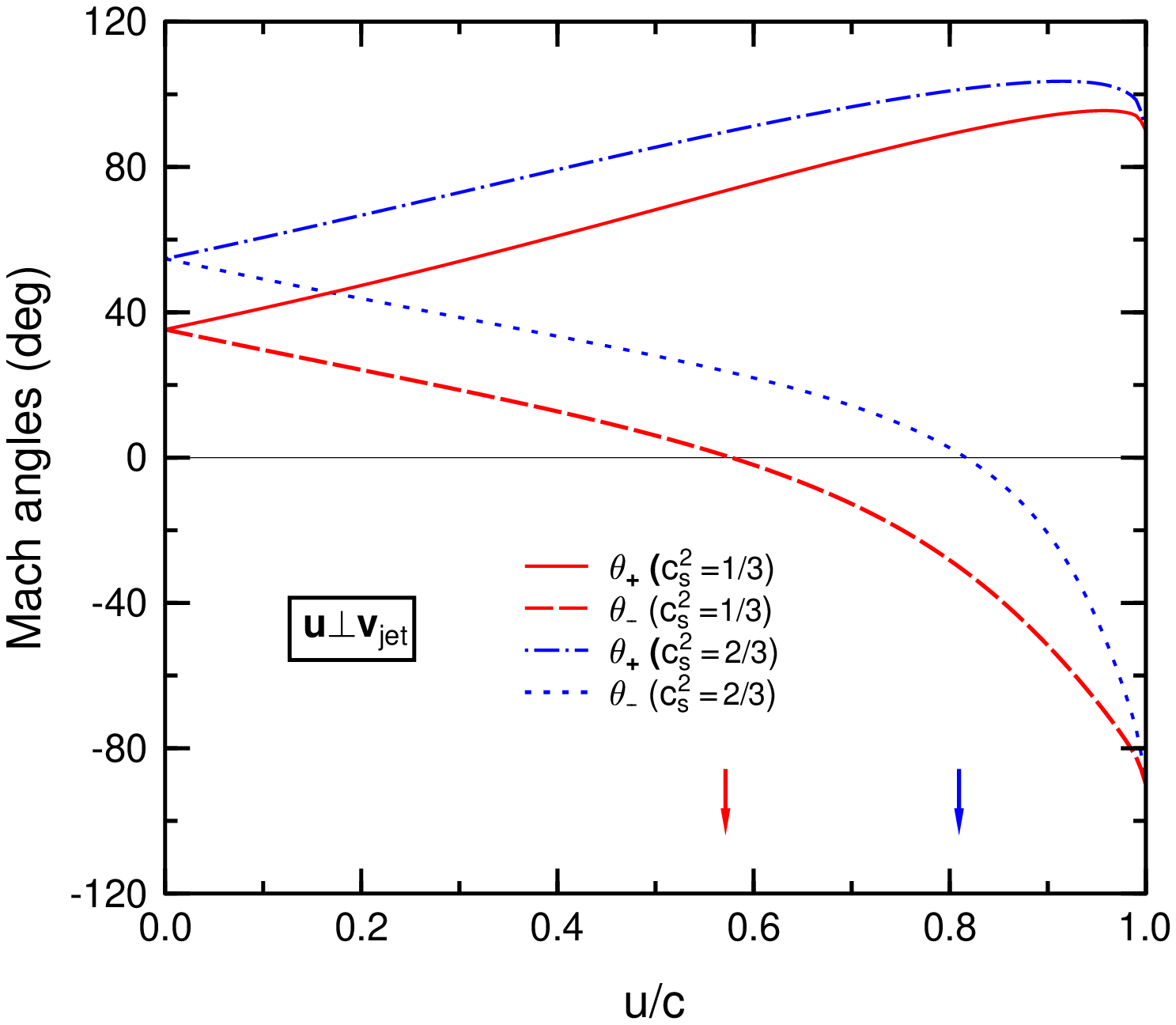}}
\caption{Angles of Mach region created by a jet moving transversely
to the fluid velocity $\bm{u}$ in the CMF. Lower (upper) set of
curves corresponds to the case $c_s^2=1/3\,(2/3)$. Arrows
mark the values $u=c_s$\,.}
\label{mach-fig2}
\end{figure}
Figure~\ref{mach-fig2} shows the numerical values of the
Mach angles for an ultrarelativistic jet moving through the QGP
transversely to its flow velocity. The solid and dashed curves are
calculated by using Eq.~(\ref{mac2}) with $c_s=\sqrt{1/3}$\,. The
dashed--dotted and dotted lines represent the corresponding angles for
$c_s=\sqrt{2/3}$\,.

\begin{figure}[hpt!]
\vspace*{-6cm}
\centerline{\epsfxsize=10cm\epsfbox{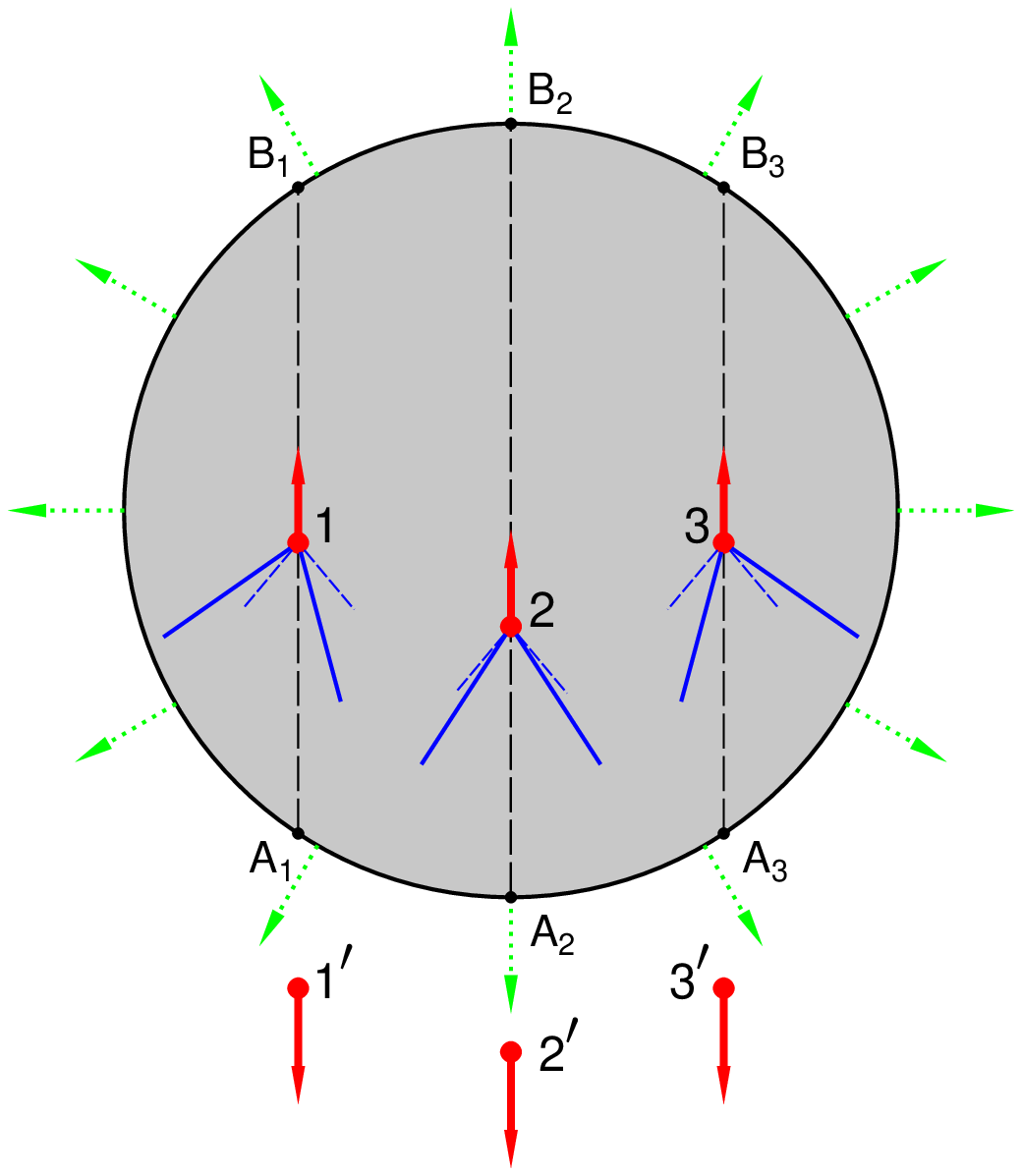}}
\caption{Schematic picture of Mach shocks from jets $1,2,3$ propagating
through the fireball matter (shaded circle) created in a central
heavy--ion collision.  Dotted arrows represent local velocities of
fireball expansion.  Thick downward arrows show associated trigger
jets. The Mach shock boundaries are shown by solid lines. Short--dashed
lines give positions of shock fronts in the case of static fireball.}
\label{mach-fig3}
\end{figure}
To discuss possible observable effects, below we consider
properties of the Mach shock created by high--energy jets propagating
in a cylindrical volume (fireball) of the QGP expanding
in radial directions. For simplicity we consider the case when
both trigger and away--side jets have zero pseudorapidities in CMF.
Presumably, this picture corresponds to the most central collisions
of equal nuclei. In Fig.~\ref{mach-fig3}
we schematically show events with different positions of di--jet
axes $A_iB_i\,(i=1,2,3)$\, with respect to the center of the
fireball. In the $2-2^\prime$ event, the away--side jet '2' propagates
along the
diameter $A_2B_2$\,, i.e. collinearly with respect to the collective
flow. In the two other cases, the di--jet axes are oriented along the chords,
$A_1B_1$ and $A_3B_3$\,, respectively. In such events, the fluid
velocity has both transverse and collinear components with respect to
the jet axis. In Fig.~\ref{mach-fig3} we also show how the
Mach fronts will be deformed in expanding matter.

The radial expansion of the fireball gives rise to the shift
of sideward peaks of the $\Delta\phi$ distributions.
This leads to an additional broadening of the
away--side maxima of the two--particle correlation function.
On average over events with different position of di--jet axes,
the peaks will be distributed in the angular interval\cite{Sat05}
\begin{eqnarray}\label{anwi}
\delta\phi\simeq\left<\theta_+ (|u_x|)-\theta_- (|u_x|)\right>
\simeq\theta_+ (\langle u_x\rangle)-\theta_- (\langle u_x\rangle)\,.
\end{eqnarray}
In the second equality a linear dependence of $\theta_\pm$ on the transverse
component of flow velocity $|u_x|$ was assumed.
Using the results of Fig.~\ref{mach-fig2} for the case $c_s2=1/3$ we
get the azimuthal distance between the peaks of about 110$^\circ$ and
the angular spread of emitted hadrons, $\delta\phi$\,, in the
range~\mbox{$25^\circ-50^\circ$} for $\langle u_x\rangle=0.2-0.4$\,.  This agrees
quite well with the positions and widths of the sideward peaks observed
by the STAR and PHENIX collaborations~\cite{Adl03a,Adl03b,Wan04,Jac05}.
On the other hand, the choice $c_s^2=2/3$ gives approximately the same
values of $\delta\phi$\,, but the predicted angular difference between two peaks
is too small, only about $70^\circ$. On the basis of this analysis we
conclude that in individual events the sideward maxima should be
located asymmetrically with respect to $\Delta\phi=\pi$ and they will be 
narrower than in an ensemble of different events.
These effects can be observed by measuring
three--particle correlations.

\vspace*{0.75cm}
The STAR-collaboration reports to see no effect of a Mach-Cone in three particle correlations:
is this due to the deflection of the jets due to tsunami-like flow of the underlying exploding plasma phase?

Distinctive features of conical flow are studied in present data with different $p_T$
windows. How can there be a perfect fluid, ideal hydrodynamic flow, but no
sound waves? They must be there, if the matter is ideal fluid-like!

PHENIX reports exactly the opposite finding of STAR: strong signs of Mach-Cones 
observed in the three particle correlations at the awayside jet!

The deformations of quark-gluon- Mach-Cones by underlying collective flow due to the high Pressure, $P_{QCD}$,
of the QGP follows the following four steps:
\begin{itemize}
\item[1.] Collective flow of Plasma yields 
\item[2.] Deformation of Mach-Cone leading to
\item[3.] Deflection of it's axis in the flow direction, resulting in
\item[4.] Away-side angular satellites being skewed and broadened
\end{itemize}

The Mach shock cones are shifted in the expanding QGP fluid. These
shifts could be directly visible in 3-particle correlations, and in the
broadening of the away-side maxima.

\section{LXD-BHs in Pb+Pb-collisions at Colliders}
The geometrical increase of the (event-by-event) LXD-BH-production probability for heavy ion collision can be as large as tenthousandfold
at the LHC, due to impact parameter - and participant number increase. Hence, the interaction of the LXD-BHs, their remnants and their
secondary Hawking-Radiation with the primary medium (quark-gluon plasma) generated by the bulk of the softer QCD interactions in the Pb+Pb
collisions may be probed: In particular, secondary Mach shock waves caused by the Hawking Monojets, secondary hydrodynamic explosions of
the Heckler-Kapusta-Hawking Plasma inside the "soft" QGP of the ion-ion collision, and the propagation of these density waves through the
Plasma deserve further studies\cite{Betz}.

\begin{figure*}[htb!]
\centerline{\includegraphics[width=0.7\textwidth]{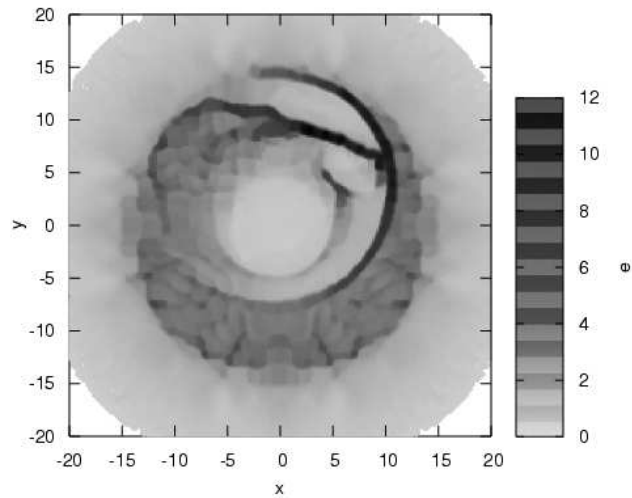}}
\caption{Mach Shock "cones" left by "slowly" moving source of high energy density "secondary" Plasma, generated by either stopped QCD-Jets
of by a secondary Heckler-Kapusta-Hawking Plasma surrounding a Black Hole created in Pb+Pb ($\sqrt{s}=5.5$ TeV) collisions, 
K. Paech, priv. communication}
\end{figure*}

\section{Conclusion}
The LHC will provide exciting discovery potential way beyond supersymmetric extensions of the SM!

\section*{Acknowledgements}
This work has been supported by FIAS, GSI, BMBF, DFG, DAAD, DOE, RIKEN, by the Alexander von Humboldt-Stiftung,
and by the ALICE collaboration.

Discussions with and important contributions by the following colleagues are acknowledged:

Barbara Betz, Leonid Satarov, Igor Mishustin, Harry Appelsh\"auser, Peter Braun-Munzinger, 
Adrian Dumitru, Ben Koch, Marcus Bleicher, Sabine Hossenfelder, Carsten Greiner, 
Tom Humanic, Kerstin Paech, Dirk Rischke, Mike Strickland, Wolfgang Cassing, Walter Greiner, 
Yasushi Nara, Daniel Henkel, Vesa Ruuskanen, Kari Eskola, Ed Shuryak, Johanna Stachel.

\end{document}